\documentclass[12pt,preprint]{aastex} 
\shorttitle{Bullet Cluster: A Challenge to $\Lambda$CDM}
\shortauthors{Lee \& Komatsu}
\begin{document}
\title{Bullet Cluster: A Challenge to $\Lambda$CDM Cosmology}
\author{Jounghun Lee}
\affil{Department of Physics and Astronomy, FPRD, Seoul National University, 
Seoul 151-747, Korea: jounghun@astro.snu.ac.kr} 
\author{Eiichiro Komatsu}
\affil{Texas Cosmology Center and Department of Astronomy, The University of
Texas at Austin,  
1 University Station, C1400 Austin, TX 78712, USA}
\begin{abstract}
To quantify how rare the bullet-cluster-like high-velocity merging
systems are in the standard $\Lambda$ Cold Dark Matter (CDM) cosmology,
we use a large-volume ($27~h^{-3}~{\rm Gpc}^3$) cosmological $N$-body 
MICE simulation to calculate the distribution of infall 
velocities of subclusters around massive main clusters. The infall-velocity 
distribution is given at $(1-3)R_{200}$ of the main cluster (where $R_{200}$ 
is similar to the virial radius), and thus it gives the distribution of 
realistic initial velocities of subclusters just before collision. 
These velocities can be compared with the initial velocities used by the 
non-cosmological hydrodynamical simulations of  1E0657-56 in the literature. 
The latest parameter search carried out by Mastropietro and Burkert have 
shown that the initial velocity of 3000~km~s$^{-1}$ at about $2R_{200}$ is 
required to explain the observed shock velocity, X-ray brightness ratio of 
the main and subcluster, X-ray morphology of the main cluster, 
and displacement of the X-ray peaks from the mass peaks. 
We show that such a high infall velocity at $2R_{200}$ is incompatible with 
the prediction of a $\Lambda$CDM model: the probability of finding 
3000~km~s$^{-1}$ in $(2-3)R_{200}$ is between $3.3\times 10^{-11}$ and 
$3.6\times 10^{-9}$. A lower velocity, 2000~km~s$^{-1}$ at $2R_{200}$, is 
also rare, and moreover,  Mastropietro and Burkert have shown 
that such a lower initial velocity does not reproduce the X-ray brightness 
ratio of the main and subcluster or morphology of the main cluster. 
Therefore, we conclude that the existence of 1E0657-56 is incompatible with 
the prediction of a $\Lambda$CDM model, unless a lower infall velocity 
solution for 1E0657-56 with $\lesssim 1800~{\rm km~s^{-1}}$ at $2R_{200}$ 
is found. 
\end{abstract}
\section{INTRODUCTION}

The bow shock in the merging cluster 1E0657-57 (also known as 
the ``bullet cluster'') observed by {\sl Chandra} indicates that the 
subcluster \citep[found by][]{Barrena-etal02} moving through this massive 
($10^{15}~ h^{-1}M_\sun$) main cluster creates a shock, and the shock
velocity is as high as $4700~ {\rm km~s^{-1}}$
\citep{Markevitch-etal02,Markevitch06}.
A significant offset between the distribution of X-ray emission and the
mass distribution has been observed \citep{Clowe-etal04,Clowe-etal06},
also indicating a high-velocity merger with gas stripped by ram pressure.

Several groups have carried out detailed investigations of the physical
properties of 1E0657-57 using non-cosmological hydrodynamical simulations
\citep{Takizawa05,Takizawa06,Milos-etal07,SF07,MB08}.
One of the key input parameters for all of these simulations is the {\it
initial velocity} of the subcluster, which is usually given at somewhere
near the virial radius of the main cluster. 

An interesting question is whether the existence of such a
high-velocity merging system is expected in a $\Lambda$CDM universe.
\citet{HW06} were the first to calculate the likelihood of 
subcluster velocities using  the Millennium Run simulation
\citep{Springel-etal05}. As the volume of the Millennium Run simulation is
limited to $(0.5~h^{-1}~{\rm Gpc})^3$, there are only 5 cluster-size halos with
$M_{200}>10^{15}~h^{-1}~M_\sun$, and 1 cluster with $M_{200}>2\times
10^{15}~h^{-1}~M_\sun$ at $z=0.28$ (close to the redshift of
1E0657-57, $z=0.296$). Therefore, \citet{HW06} had to extrapolate their
results for $M_{200}>10^{14}~h^{-1}~M_\sun$ assuming that the
likelihood of finding the bullet-cluster systems scales with
$V_{\rm sub}/V_{200}$, where $V_{\rm sub}$ is the subcluster velocity in
the rest frame of the main cluster, and $V_{200}=(GM_{200}/R_{200})^{1/2}$. 
Here, $R_{200}$ is the radius within which the mean mass density is 200 
times the critical density of the universe, and $M_{200}$ is the mass 
enclosed within $R_{200}$.

While \citet{HW06} concluded that the existence of 1E0657-57 is
consistent with the standard $\Lambda$CDM cosmology, this conclusion was
later challenged by \citet{FR07} who showed that, once an updated mass of 
the main cluster of 1E0657-57 is taken into account, the probability of 
finding 1E0657-57 is as low as $10^{-7}$. This conclusion still relies on 
the extrapolation of the likelihood derived for 
$M_{200}>10^{14}~h^{-1}~M_\sun$. 

As the probability of finding high-velocity merging systems decreases
exponentially with velocities, an accurate determination of the
{\it subcluster velocity}, rather than the shock velocity, is crucial. 
\citet{Milos-etal07} and \citet{SF07} used hydrodynamical simulations to
show that the subcluster velocity can be significantly lower than the
shock velocity (which is $4700~{\rm km~s^{-1}}$). 
\citet{Milos-etal07} found that the subcluster velocity 
can be $4050~{\rm km~s^{-1}}$, whereas \citet{SF07} found 
that it can be as low as $2700~{\rm km~s^{-1}}$.
\citet{MB08} showed that the subcluster velocity of $3100~{\rm
km~s^{-1}}$ best reproduces the X-ray data of 1E0657-57.

These varying results are due in part to the varying assumptions about
the initial velocity given to the subcluster at the beginning of their
hydrodynamical simulations: 
\citet{Milos-etal07} used zero relative velocity between the main cluster
and subcluster at the initial separation of 4.6~Mpc (which is 2 times
$R_{200}$ of the main cluster, 2.3~Mpc). The velocity is about
1600~km~s$^{-1}$ at a separation of 3.5~Mpc ($\simeq
1.5R_{200}$)\footnote{Milosavljevi\'c (2010), private 
communication. All velocities quoted throughout this paper are
calculated in the rest frame of the main cluster.};
\citet{SF07} used the initial velocity of 2057~km~s$^{-1}$ when the
separation was 3.37~Mpc ($\simeq 1.5R_{200}$); and \citet{MB08}
explored various initial velocities such as 2057~km~s$^{-1}$ at the initial
separation of 3.37~Mpc and 2000, 3000, and 5000~km~s$^{-1}$ at the initial
separation of 5~Mpc ($\simeq 2.2R_{200}$). \citet{MB08} found that the
simulation run with the initial velocity of 3000~km~s$^{-1}$ best reproduces 
the X-ray data.

In this paper, we demonstrate that the initial velocities used by 
\citet{Milos-etal07} and \citet{SF07} are consistent with the prediction
of a $\Lambda$CDM model, but those of \citet{MB08} at 5~Mpc are
not. The simulations of \citet{Milos-etal07} and \citet{SF07} do not
reproduce details of the X-ray and weak lensing data of 1E0657-57, and
\citet{MB08} argue that one needs the initial velocity of
3000~km~s$^{-1}$ to explain the data. If this is true, the existence of
1E0657-57 is incompatible with the prediction of a $\Lambda$CDM model. 

\section{FINDING CLUSTERS OF CLUSTERS IN SIMULATION}

As high-velocity systems are rare, it is crucial to use a large-volume
simulation to derive the reliable probability distribution. The previous
study is somewhat inconclusive due to the limited volume of the
Millennium Run simulation, $(0.5~h^{-1}~{\rm Gpc})^3$.
We calculate the probability of finding bullet-like
systems using a simulation with substantially larger volume,
$(3~h^{-1}~{\rm Gpc})^3$. 

We use the publicly available simulated dark-matter halo catalogs at $z=0$ and
$0.5$, which are constructed from the largest-volume $N$-body
Marenostrum Institut de Ci\`encies de l'Espai(MICE) simulations 
\citep{MICE09}. They used the publicly available GADGET-2 code 
\citep{Springel05}, with the cosmological parameters of 
$\Omega_{m}=0.25,\ \Omega_{\Lambda}=0.75,\ 
\Omega_{b}=0.044,\ h=0.7,\ n_{s}=0.95$, and $\sigma_{8}=0.8$. These
numbers are consistent with those derived from the seven-year data of the
{\sl Wilkinson Microwave Anisotropy Probe} \citep{wmap7}.

The MICE simulation that we shall use in this paper has the particle
mass of $M_{\rm par}=23.42\times 10^{10}~ h^{-1}~M_\sun$ and the linear box 
size of $L_{\rm box}=3072~ h^{-1}$~Mpc. The standard friends-of-friends (FoF)
algorithm \citep{FoF02} with the linking length parameter of $b=0.2$ was 
employed to find the cluster halos from the distribution of $2048^{3}$ 
dark matter particles. See \citet{Fosalba-etal08} and \citet{MICE09} for a 
detailed description of the MICE simulations and the halo-identification 
procedure.

The halos identified in the MICE simulation contain at least 143 $N$-body
particles. The derived halo catalog contains the center-of-mass
positions (${\bf X}$) and velocities (${\bf V}$) of halos, as well as
the number of particles in each halo ($N_{\rm par}$). 
Note that the number of particles in each halo has been corrected for
the known systematic effect of the FoF algorithm, using 
$N_{\rm par}^{\rm corr}=N_{\rm par}(1-N_{\rm par}^{-0.6})$
\citep{Warren-etal06,MICE09}.

The mass of each halo is calculated as $N_{\rm par}$ times the mass of each
particle, $M_{\rm par}$. The mass of halos identified by FoF with the
linking length of 0.2 {\it approximately} corresponds to $M_{200}$, i.e., the
mass within $R_{200}$, within which the overdensity is 200 times the
critical density of the universe at a given redshift,
$M_{200}=\frac{4\pi}{3}[200\rho_c(z)]R_{200}^3$. 
It is, however, known that the FoF mass tends to be larger than
$M_{200}$, especially for high-mass clusters which are less
concentrated \citep{Lukic-etal09}. As a result, $R_{200}$ we quote in
this paper may be an overestimate. 

The difference between the FoF mass and $M_{200}$ decreases as the
number of particles per halo, $N_{\rm par}$, increases \citep{Lukic-etal09}. 
For the main halo masses of our interest, $M_{\rm main}\ge 0.5\times
10^{15}~h^{-1}~M_\sun$, the average value of $N_{\rm par}$ is 
3355 and 3160 at $z=0$ and $0.5$, respectively. Using this, we estimate
that, on average, our $R_{200}$ may be 10\% too large. 
This error is insignificant for our purpose. Moreover, as correcting this error
strengthens our conclusion by making the probability of finding
high-velocity subclusters even smaller, we shall ignore the difference
between $R_{200}$ and the radius estimated from the FoF mass.

To find the ``clusters of clusters''
(i.e., groups of clusters with one massive main cluster 
surrounded by many less massive satellite clusters), we treat 
each cluster in the catalog as a particle and re-apply the FoF algorithm with 
the linking length of 0.2. This time, the linking length of 0.2 means
the length of 0.2 times $L_{\rm box}/(N_{\rm cl})^{1/3}$, where $N_{\rm
cl}$ is the total number of clusters found in the simulation (2.8 and 1.7 
million clusters at $z=0$ and $0.5$, respectively).
Each cluster of clusters has the ``main cluster,'' or the most massive
member of each cluster of clusters. All the other clusters are called
``satellite clusters'' or ``subclusters.'' 
Table \ref{tab:fof} shows the total number of cluster-size halos found in the
simulation, the number of clusters of clusters having at least two
members, and the mean mass of main clusters.
For each main cluster, we calculate $R_{200}$ from its mass as 
$R_{200}=[3M_{200}/(4\pi\times 200\rho_{c}(z))]^{1/3}$. Most of the
satellite clusters are located at $r\gtrsim 2R_{200}$ from the main
cluster, where $r$ is the distance between the main cluster and its
satellites. 

\section{DERIVING THE INFALL VELOCITY DISTRIBUTION}

Our goal in this paper is to derive the distribution of infall
velocities around the main clusters. To compare with the initial
velocities used by the hydrodynamical simulations in the literature
\citep{Milos-etal07,SF07,MB08}, we calculate the infall velocity
distribution within $(2-3)R_{200}$ \citep{MB08}, at $1.5R_{200}$
\citep{Milos-etal07,SF07}, and at $R_{200}$.

We define the pairwise velocity of a satellite cluster, ${\bf V}_{c}$,
as the velocity of the satellite relative to that of the main cluster,  
${\bf V}_{c}\equiv {\bf V}_{\rm main}-{\bf V}_{\rm sat}$. 
When satellite clusters are close to the main cluster, $V_c$ must be
strongly influenced (if not completely determined) by the gravitational
potential of the main cluster. Thus, $V_c$ should depend on the main
cluster mass, $M_{\rm main}$. If $V_c$ is solely determined by the
gravitational potential of the main cluster, then $V_c\propto M_{\rm
main}^{1/2}$. In reality, however, it is not only the gravity of the
main cluster but also the influences from the surrounding large-scale 
structures that should determine $V_c$ \citep{benson05,wang-etal05,wetzel10}. 

\begin{figure}[ht]
\begin{center}
\plotone{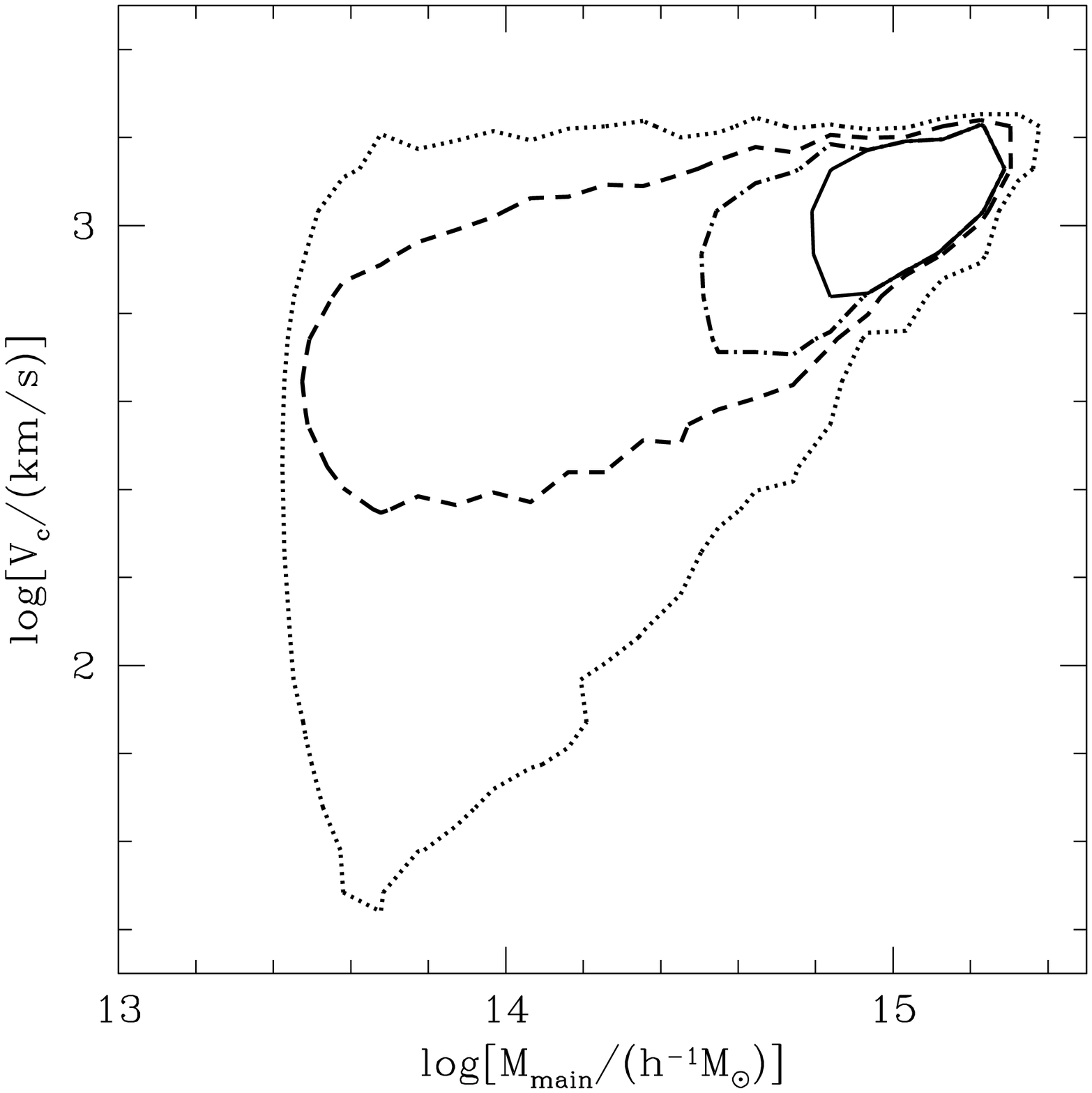}
\caption{Distribution of satellite clusters in the $V_{c}$-$M_{\rm
 main}$ plane. The dotted line shows all the satellite clusters found in
 the simulation at $z=0$; the dashed line shows those lying between 
$2R_{200}\le r\le 3R_{200}$ from the main cluster; the dot-dashed line
 shows those lying between $2R_{200}\le r\le 3R_{200}$ and about to undergo
 nearly head-on collisions with $\vert\cos\theta\vert \ge 0.9$; and the
 solid line shows those lying between $2R_{200}\le r\le 3R_{200}$, about
 to undergo nearly head-on collisions, and having small masses compared
 to the main cluster mass, $M_{\rm sat}/M_{\rm main}\le 1/10$, 
 where $M_{\rm main}\ge 0.7\times 10^{15}~h^{-1}~M_\sun$.}
\label{fig:contour}
\end{center}
\end{figure}

\begin{figure}
\begin{center}
\plotone{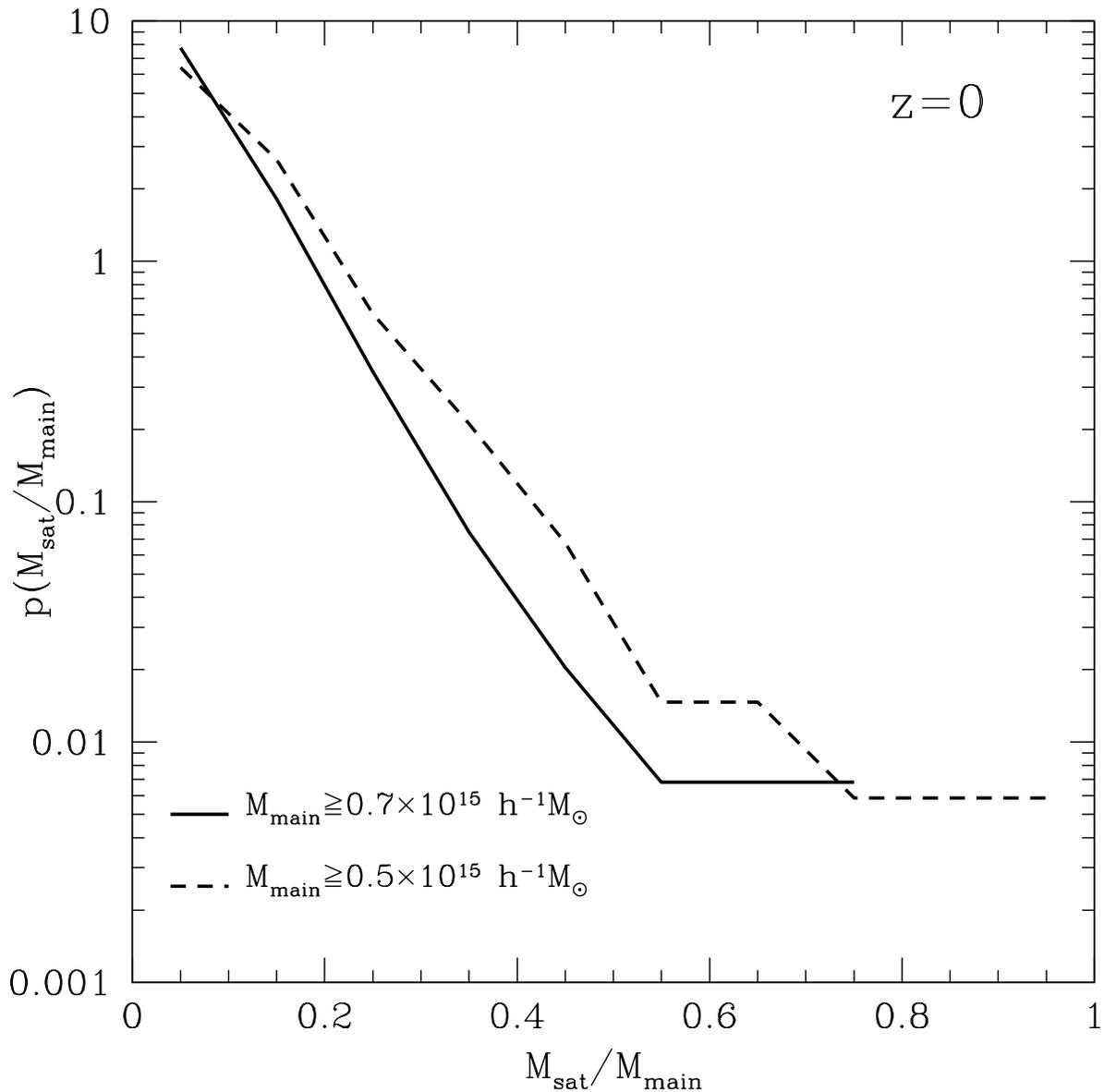}
\caption{Distribution of the sub-main cluster mass ratio, $M_{\rm
 sat}/M_{\rm main}$, at $z=0$. The solid and dashed lines show the
 distribution for the main cluster masses of $M_{\rm main}>0.7$ and
 $0.5\times 10^{15}~h^{-1}~M_\sun$, respectively. 
The distribution is normalized to unity when integrated, $\int_0^1
 p(x)dx=1$.}
\label{fig:massrat1}
\end{center}
\end{figure}

\begin{figure}
\begin{center}
\plotone{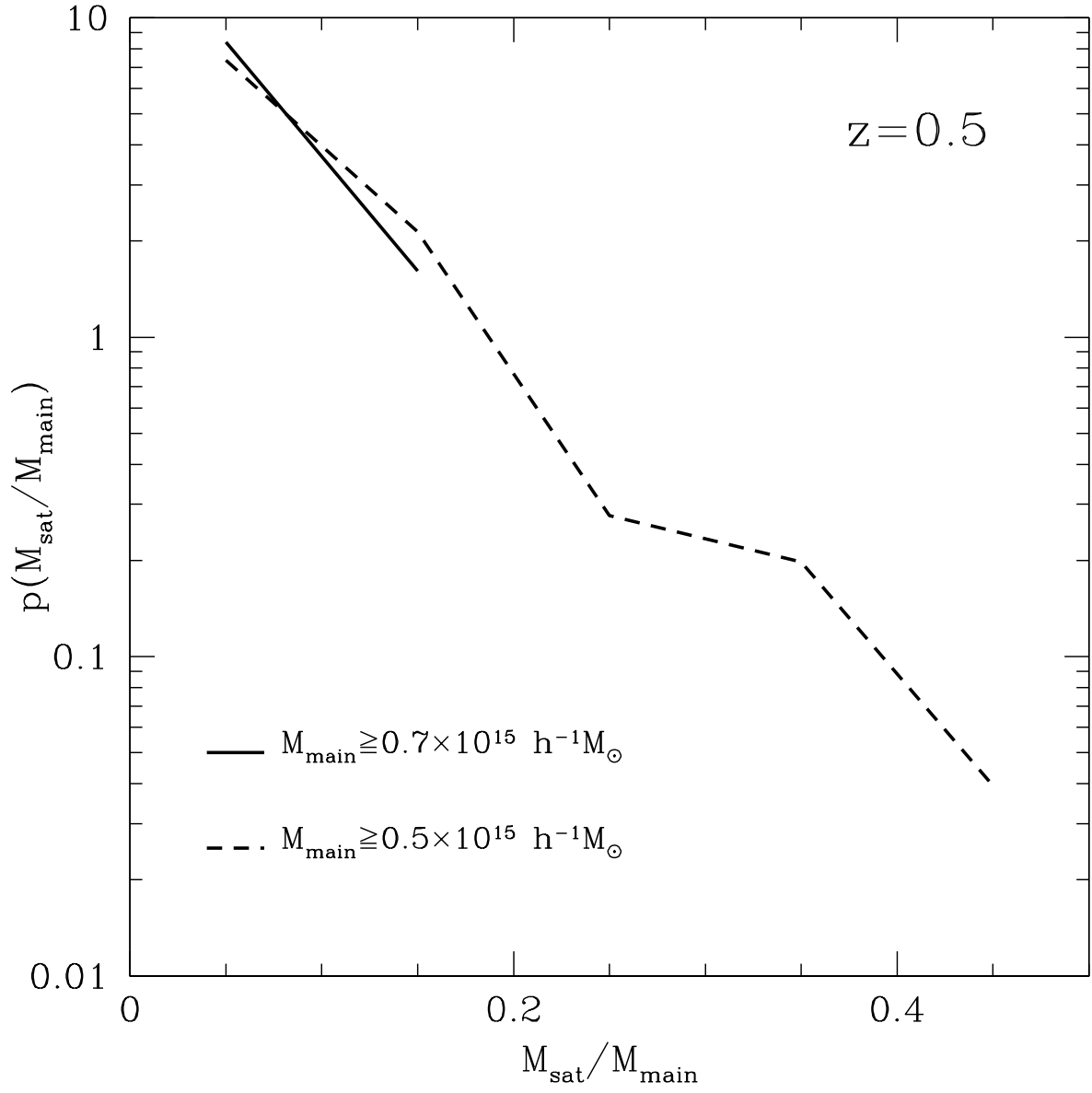}
\caption{Same as Figure \ref{fig:massrat2}, but for $z=0.5$.}
\label{fig:massrat2}
\end{center}
\end{figure}

Figure \ref{fig:contour} shows the distribution of satellite clusters in the 
$\log V_{c}-\log M_{\rm main}$ plane (dotted line) at $z=0$. 
There is a clear correlation between $V_c$ and $M_{\rm main}$ (the
larger the $M_{\rm main}$ is, the larger the $V_c$ becomes), although
it is not simply $V_c\propto M_{\rm main}^{1/2}$.
The dotted line in Figure \ref{fig:contour} shows the distribution of
all satellite clusters. Next, we shall select the satellite clusters
that belong to bullet-like systems. We define the bullet-like system as
follows: the main cluster exerts dominant gravitational force on
satellite clusters, and at least one satellite cluster is on its way of
head-on merging with the main cluster. More specifically, the following
three criteria are used to select the candidate bullet cluster systems
from the clusters of clusters at a given $z$:
\begin{enumerate}
\item[1.] Satellite clusters lie between $2R_{200}\le r\le 3R_{200}$,
	   and thus their motion is predominantly determined by the
	   gravitational potential of the main cluster,
\item[2.] Satellite clusters are about to undergo nearly head-on
	    collisions with the main cluster: $\vert{\bf V}_{c}\cdot{\bf
 r}\vert/(\vert{\bf V}_{c}\vert\vert{\bf  r}\vert)\ge 0.9$, and 
\item[3.] The mass of satellites is less than or equal to 10\% of that of the
	     main cluster, $M_{\rm sat}/M_{\rm main}\le 1/10$, and the
	  main cluster mass is greater than some value, $M_{\rm main}\ge
	  M_{\rm crit}$.
\end{enumerate}
The third criterion is motivated by the observation of 
1E0657-57 indicating that the mass of the bullet subcluster is an
order-of-magnitude lower than that of the massive main cluster, and the
mass of the main cluster is $\sim 10^{15}~h^{-1}~M_\sun$ \citep{SF07}.
As the latest simulation by \citet{MB08} showed that the mass ratio of
$6:1$ best reproduces the observed data of 1E0657-56 
\citep[also see][]{nusser08}, we have also studied the case 
with 
$M_{\rm sat}/M_{\rm main}\le 1/5$, finding similar results; thus, our 
conclusion is insensitive to the precise value of the mass ratio.
In Figure \ref{fig:massrat1} and \ref{fig:massrat2}, we show the
distribution of the mass ratio, $M_{\rm sat}/M_{\rm main}$, at $z=0$ and
$0.5$, respectively. As expected, larger-$M_{\rm sat}/M_{\rm main}$
(i.e., closer-to-major-merger) collisions are exponentially rare. 
This makes 1E0657-57 even rarer, if the mass ratio is as large as 
$M_{\rm sat}/M_{\rm main}=1/6$. For the rest of the paper, we shall study 
the case of $M_{\rm sat}/M_{\rm main}<1/10$, keeping in mind that 1E0657-57 
can be even rarer than our study indicates.
 
In Figure \ref{fig:contour}, we show the distribution of satellite
clusters satisfying the condition 1 (dashed line), 
the conditions 1 and 2 (dot-dashed line), and the conditions 1, 2, and 3
(solid line). Note that $V_c$ of the satellite clusters that satisfy all of 
the above conditions approximately follows $V_{c}\propto M^{1/2}_{\rm main}$. 
This is an expected result, as the satellite clusters in this case are 
basically point masses (nearly) freely falling into the main cluster.
We also find similar results for $z=0.5$.

In Table~\ref{tab:z0} and \ref{tab:z05}, we show the number of
bullet-like systems satisfying all of the above conditions at $z=0$ and
$0.5$, respectively. At $z=0$, about 1 in 3 clusters of clusters
with $M_{\rm main}\ge 0.7\times 10^{15}~h^{-1}~M_\sun$
contains a nearly head-on collision subcluster. At $z=0.5$, 
about 1 in 5 clusters of clusters with 
$M_{\rm main}\ge 0.7\times 10^{15}~h^{-1}~M_\sun$ contains a nearly head-on 
collision subcluster. Therefore, head-on collision systems are quite common 
- but, how about their infall velocities?  

\begin{figure}
\begin{center}
\plotone{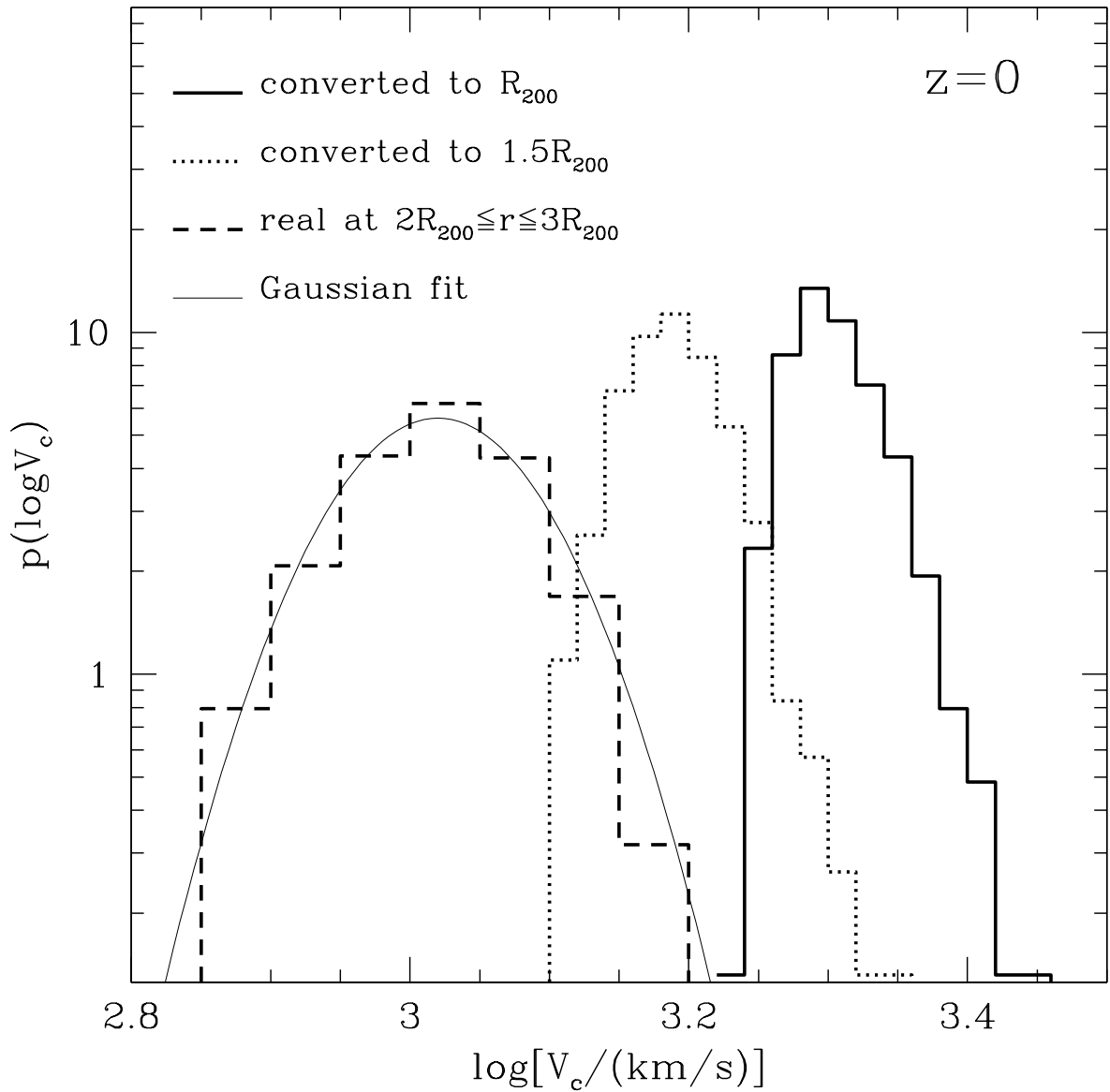}
\caption{Probability density distribution of the infall velocities, 
$\log V_c$, of the bullet-cluster-like systems at $z=0$. The main cluster 
masses are $M_{\rm main}\ge 0.7\times 10^{15}~h^{-1}~M_\sun$, for which there 
are 1135 bullet-like systems in the simulation at $z=0$. The dashed line 
shows the distribution of $\log V_c$ within $2\le r/R_{200}\le 3$ measured 
from the simulation. This distribution shows that the initial velocities used 
by \citet{MB08}, $V_c\ge 2000~{\rm km~s^{-1}}$ at $2.2R_{200}$, are 
incompatible with the prediction of a $\Lambda$CDM model: {\it none} 
(out of 1135 eligible samples) has the velocity as high as 
$V_c\ge  2000~{\rm km~s^{-1}}$ in $2\le r/R_{200}\le 3$. The dotted and solid 
lines show the distribution of $V_c$ at $1.5R_{200}$ and $R_{200}$, 
respectively,which are obtained by converting the dashed line using equation
(\ref{eqn:vc}). We also show a Gaussian fit to the dashed line, which
is given by equation~(\ref{eqn:fit}).
Note that $10^{3.2}=1585$, $10^{3.3}=1995$, and $10^{3.4}=2512$.}
\label{fig:pro1}
\end{center}
\end{figure}

\begin{figure}
\begin{center}
\plotone{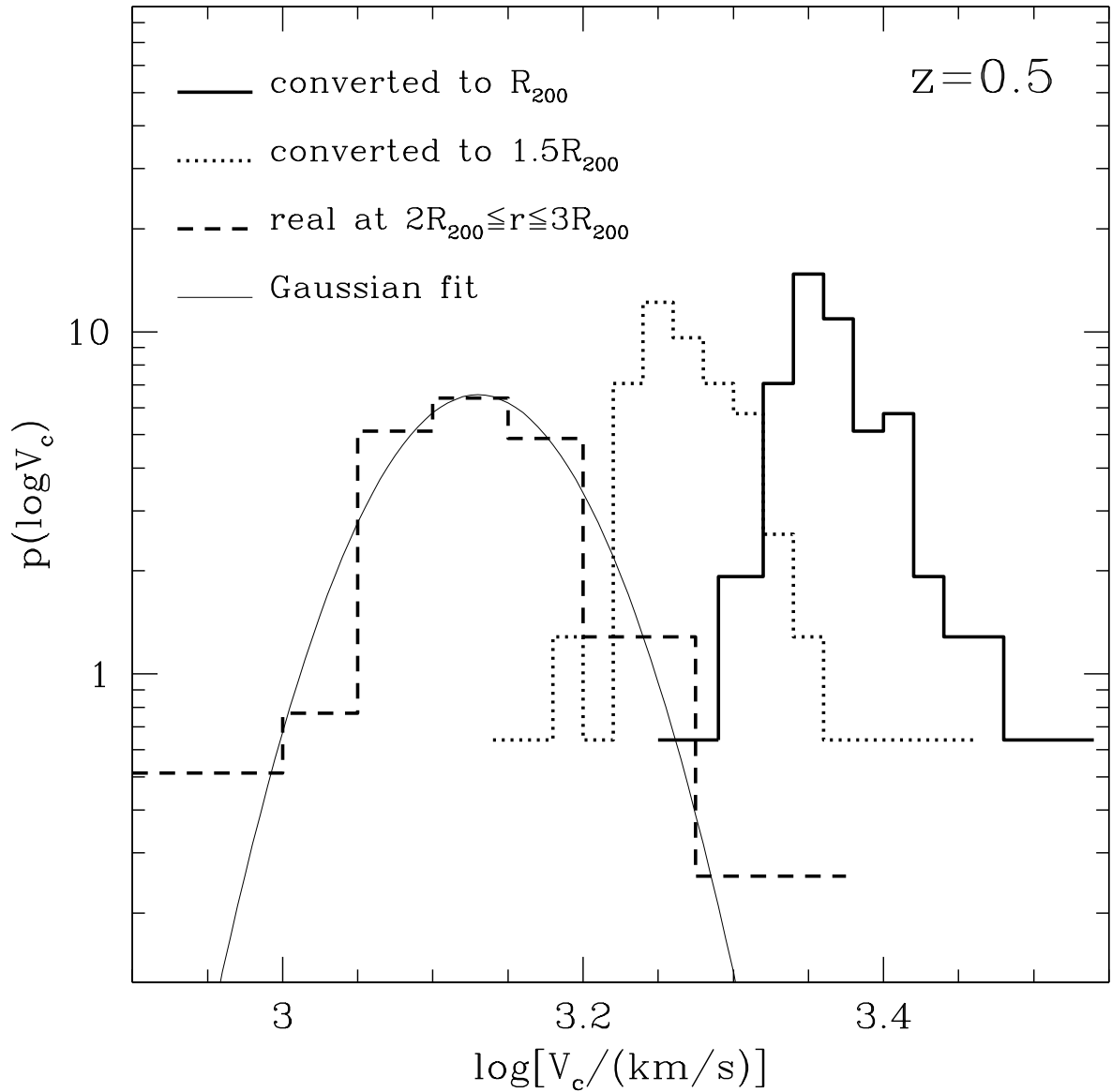}
\caption{Same as Figure \ref{fig:pro1}, but for $z=0.5$.
The main cluster masses are $M_{\rm main}\ge 0.7\times 
10^{15}~h^{-1}~M_\sun$, for which there are 177 bullet-like systems in the 
simulation at $z=0.5$. None has the velocity as high as 
$V_c=3000~{\rm km~s^{-1}}$ in $2\le r/R_{200}\le 3$, while there is one 
subcluster with the velocity of $V_c\ge 2000~{\rm km~s^{-1}}$ in 
$2\le r/R_{200}\le 3$. (This subcluster has $V_c=2049~{\rm km~s^{-1}}$.) 
We also show a Gaussian fit to the dashed line, which is given by 
equation~(\ref{eqn:fit}). Note that 
$10^{3.2}=1585$, $10^{3.3}=1995$, and $10^{3.4}=2512$.}
\label{fig:pro2}
\end{center}
\end{figure}

We calculate the probability density distribution of $\log V_{c}$ using
the selected bullet-like systems (within $(2-3)R_{200}$) at $z=0$ and
$0.5$. The results for $M_{\rm main}\ge 0.7\times 10^{15}~h^{-1}~M_\sun$ 
are shown in Figure~\ref{fig:pro1} ($z=0$) and \ref{fig:pro2} ($z=0.5$).
A striking result seen from Figure~\ref{fig:pro1} is that, of 1135
bullet-like systems shown here for $z=0$, {\it none} has the infall
velocity as high as 3000~km~s$^{-1}$, which is required to
explain the X-ray and weak lensing data of 1E0657-56 \citep{MB08}.
A lower velocity, 2000~km~s$^{-1}$, is also rare: 
{\it none} (out of 1135) within $(2-3)R_{200}$ has $V_c\ge
2000~{\rm km~s~^{-1}}$ at $z=0$.

We find a similar result for $z=0.5$ (Figure~\ref{fig:pro2}): {\it none}
(out of 78) has the infall velocity as high as 3000~km~s$^{-1}$, and only 
one has $V_c\ge 2000~{\rm km~s~^{-1}}$. However, we would need better 
statistics (i.e., a bigger simulation) at $z=0.5$ to obtain more accurate 
probability. In any case, \citet{MB08} argued that an infall velocity of
2000~km~s$^{-1}$ is not enough to explain the X-ray brightness ratio of
the main and subcluster or the X-ray morphology of the main
cluster. These results indicate that the existence of 
1E0657-56 rules out $\Lambda$CDM, unless a lower infall velocity
solution for 1E0657-56 is found. 

The significance increases if we lower the minimum main cluster mass.
\citet{MB08} argue that $M_{\rm main}\sim 0.5\times
10^{15}~h^{-1}~M_\sun$ fits the data of 1E0657-56 better. 
For a lower minimum main cluster mass, $M_{\rm main}\ge 0.5\times
10^{15}~h^{-1}~M_\sun$, none out of 2189 bullet-like systems at $z=0$
has $V_c\ge 2000~{\rm km~s~^{-1}}$, 
none out of 186 systems at $z=0.5$ has $V_c\ge 3000~{\rm km~s~^{-1}}$,
and only one system at $z=0.5$ has $V_c\ge 2000~{\rm km~s~^{-1}}$.

To examine whether or not the above results depend on the value 
of the linking length parameter, $b$, of the FoF algorithm used for
finding clusters of clusters, we have repeated 
all the analyses by varying the values of $b$ from 0.15 to 0.5. We have
found similar results at both redshifts, 
demonstrating that our conclusion is insensitive to the exact values of
$b$ used for the identification of clusters of clusters with the FoF
algorithm.

To compare with the initial velocities used by the other simulations 
\citep{Milos-etal07,SF07}, we need to calculate the infall velocity
distribution at $1.5R_{200}$. As most of the subclusters  are located at
$r\gtrsim 2R_{200}$, we have much fewer subclusters in
$(1-2)R_{200}$. (There are only 191 subclusters within $(1-2)R_{200}$ at
$z=0$.) To solve this problem and keep the good statistics, we
shall use the following simple dynamical model to convert the results
in $(2-3)R_{200}$ to those at $1.5R_{200}$ as well as at $R_{200}$.

The motion of the subclusters located in $(2-3)R_{200}$ is predominantly
determined by the gravitational potential of the main halo. This is
especially true for those in a nearly head-on collision course (i.e.,
nearly a radial orbit); thus, one may treat a selected sub-main
cluster system as an isolated two-body system.  
Under this assumption, the pairwise velocity at $r_{\rm in}<2R_{200}$ is
given in terms of the velocity at $r_{\rm out}\ge 2R_{200}$ (which is
measured from the simulation) and the mass of the main halo (which is
also measured from the simulation):  
\begin{equation}
\label{eqn:vc}
V^2_{c}(r_{\rm in})=
V_{c}^{2}(r_{\rm out}) + 
\frac{2GM_{\rm main}}{R_{200}}
\left(
\frac{R_{200}}{r_{\rm in}}-\frac{R_{200}}{r_{\rm out}}
\right),
\end{equation}
where $G= 4.3\times 10^{-9}~{\rm km^2~s^{-2}}~M^{-1}_\sun~{\rm Mpc}$ is 
Newton's gravitational constant.  

In Figure~\ref{fig:pro1} and \ref{fig:pro2}, we show 
the probability density distribution of $\log V_{c}$ at $z=0$ and $0.5$, 
respectively. The dashed lines show the original distribution for 
$(2-3)R_{200}$, while the dotted and solid lines show the distribution at 
$1.5R_{200}$ and $R_{200}$, respectively, computed from 
equation~(\ref{eqn:vc}).
We find that the initial velocities used by \citet{Milos-etal07} 
($\approx 1600$~km~s$^{-1}$) and  \citet{SF07} ($\approx 2000$~km~s$^{-1}$) 
are consistent with the predictions of a $\Lambda$CDM model: at $1.5R_{200}$, 
9 (out of 1135) subclusters have $V_c\ge 2000$~km~s$^{-1}$ at $z=0$, and 16 
(out of 117) subclusters have $V_c\ge 2000$~km~s$^{-1}$ at $z=0.5$.
However, these simulations do not reproduce the details of the X-ray and 
weak lensing data of 1E0567-56 \citep{MB08}, and thus this agreement does 
{\it not} imply that the existence of 1E0567-56 is 
consistent with $\Lambda$CDM.

\begin{figure}
\begin{center}
\plotone{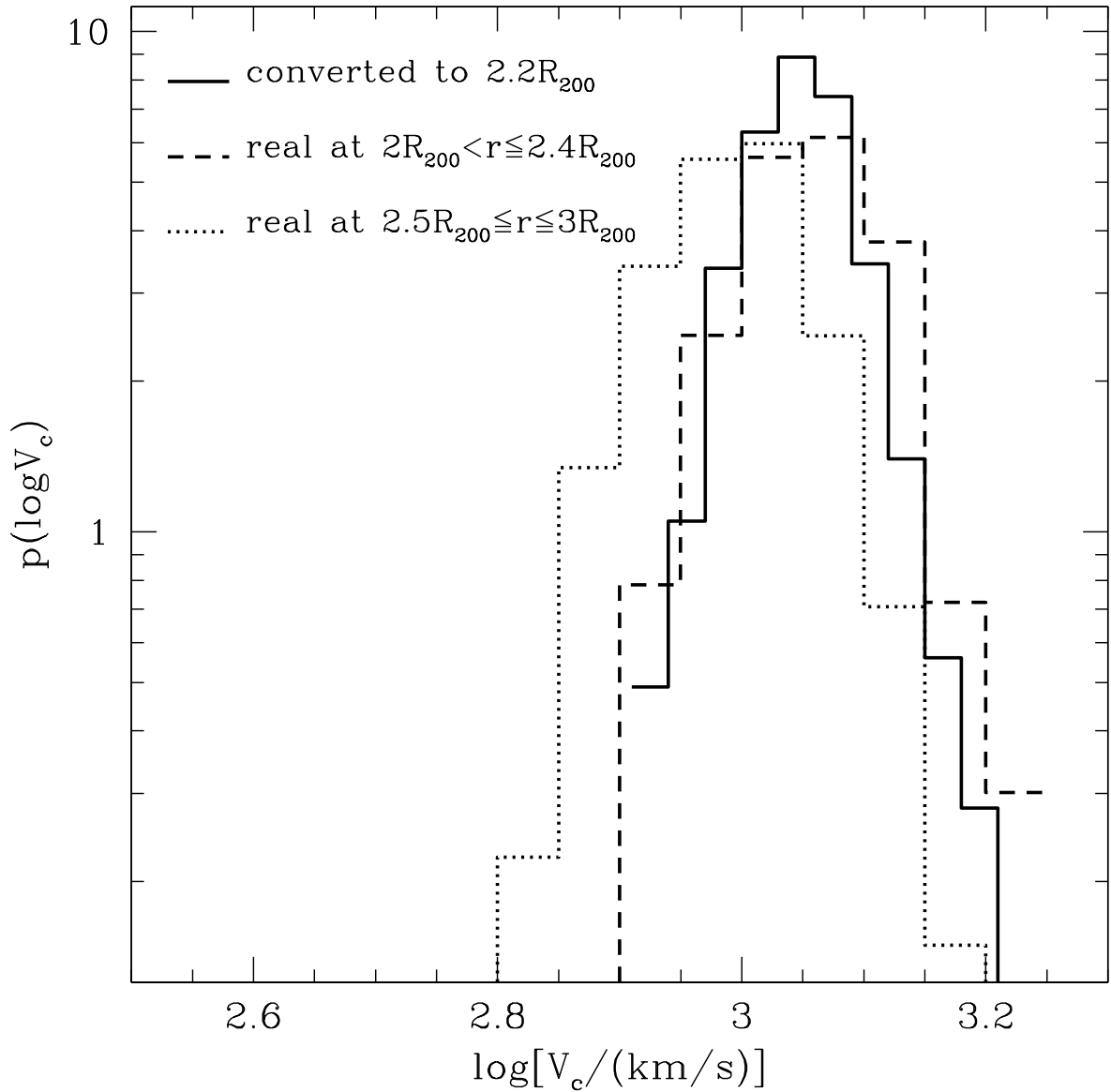}
\caption{
Testing equation~(\ref{eqn:vc}). The dashed line shows the distribution
 of $\log V_{c}$ in $2\le r/R_{200}\le 2.4$ measured from the
 simulation, while the solid line shows the distribution of $\log V_{c}$
 at $r_{\rm in}=2.2R_{200}$ calculated from the measured distribution in 
$2.5\le r_{\rm out}/R_{200}\le 3$ (dotted line) and equation~(\ref{eqn:vc}).}
\label{fig:test}
\end{center}
\end{figure}

How reliable is this extrapolation of the infall velocity? 
To check the accuracy of equation~(\ref{eqn:vc}), we compare 
$p(V_{c})$ in $2 \le r/R_{200}\le 2.4$ measured from the simulation and 
$p(V_{c})$ at $2.2R_{200}$ computed from equation~(\ref{eqn:vc}). 
Specifically, we use equation~(\ref{eqn:vc}) to calculate the velocity at 
$r_{\rm in}=2.2R_{200}$ from velocities in 
$2.5R_{200}\le r_{\rm out}\le 3R_{200}$. In Figure \ref{fig:test}, we show the 
measured $p(V_{c})$ in $2 \le r/R_{200}\le 2.4$  (dashed line), the predicted 
$p(V_{c})$ at $2.2R_{200}$ (solid line), and the original $p(V_{c})$ in
$2.5 \le r/R_{200}\le 3$  (dotted line). We find an excellent agreement
between the measured and predicted distribution. 

\section{DISCUSSION AND CONCLUSION}

\citet{MB08} showed that the subcluster initial velocity of
3000~km~s$^{-1}$ at the separation of 5~Mpc is required to explain the
X-ray and weak lensing data of 1E0657-56 at $z=0.296$. They argued that
a lower velocity, 2000~km~s$^{-1}$, seems excluded because it cannot
reproduce the observed X-ray brightness ratio of the main and
subcluster or the X-ray morphology of the main cluster. 

In this paper, we have shown that such a high velocity at 
5~Mpc, which is about 2 times $R_{200}$ of the main cluster, is
incompatible with the prediction of a $\Lambda$CDM model. 
Using the results at $z=0$ and $M_{\rm main}\ge 0.7\times
10^{15}~h^{-1}~M_\sun$, $\Lambda$CDM is excluded by more than 99.91\% 
confidence level (none out of 1135 subclusters has $V_c\ge
2000$~km~s$^{-1}$ in $2\le r/R_{200}\le 3$).
For a lower minimum main cluster mass, 
$M_{\rm main}\ge 0.5\times 10^{15}~h^{-1}~M_\sun$, 
$\Lambda$CDM is excluded by more than 99.95\% 
confidence level (none out of 2189 subclusters has $V_c\ge
2000$~km~s$^{-1}$ in $2\le r/R_{200}\le 3$).

The results at $z=0.5$ are not yet fully conclusive due to the limited 
statistics: none out of 78 subclusters has $V_c\ge
3000$~km~s$^{-1}$ in $2\le r/R_{200}\le 3$, while there is one 
subcluster with $V_c\ge 2000$~km~s$^{-1}$ in $2\le r/R_{200}\le
3$. For $M_{\rm main}\ge 0.5\times 10^{15}~h^{-1}~M_\sun$, 
none out of 186 subclusters has $V_c\ge 3000$~km~s$^{-1}$, while there is one 
subcluster with $V_c\ge 2000$~km~s$^{-1}$.

While these confidence levels are directly measured from the simulation,
one can estimate the probability better by fitting the probability density, 
$p(\log V_c)$, to a Gaussian distribution as
\begin{equation}
 p(\log V_c) = \frac{1}{\sqrt{2\pi\sigma^2_{\nu}}}
{\exp\left[-\frac{\left(\log V_c-\nu\right)^2}{2\sigma^2_{\nu}}\right]},
\label{eqn:fit}
\end{equation}
where $V_c$ is in units of km~s$^{-1}$ and $\nu$ and $\sigma_{\nu}$ are 
the two fitting parameters. The best-fit values of the two parameters 
for $z=0$ and $0.5$ are $(\nu,\sigma_\nu)=(3.02,0.07)$ and
$(3.13,0.06)$, respectively. The mean velocity at $z=0$ is smaller than that 
at $z=0.5$ by a factor of $10^{3.13-3.02}=1.29$. This may be understood as 
the effect of $\Lambda$ slowing down the structure formation at $z<0.5$.
 
Generally, one has to be careful about this approach, as we are probing
the tail of the distribution, where the above fits may not be accurate. 
Using the above Gaussian fits, we find 
$P(>3000~{\rm km~s^{-1}})=3.3\times 10^{-11}$ and 
$3.6\times 10^{-9}$ at $z=0$ and $0.5$, respectively. 
We also find $P(>2000~{\rm km~s^{-1}})=2.9\times 10^{-5}$ and 
$2.2\times 10^{-3}$ at $z=0$ and $0.5$, respectively. These numbers pose a 
serious challenge to $\Lambda$CDM, unless one finds a lower velocity solution 
for 1E0657-56. Here, a ``lower velocity'' may be somewhere between 
$V_c\lesssim 1500$ and $1800$~km~s$^{-1}$ at $r\sim 2R_{200}$, which give 
1\% probabilities at $z=0$ and $z=0.5$, respectively. 

The bullet cluster 1E0657-56 is not the only site of violent cluster
mergers. For example, there are A520 \citep{Markevitch-etal05} and 
MACS~J0025.4-1222 \citep{Bradac-etal08}. Also, high-resolution mapping 
observations of the Sunyaev-Zel'dovich (SZ) effect have revealed a violent 
merger event in RX~J1347-1145 at $z=0.45$ 
\citep{Komatsu-etal01,Kitayama-etal04,Mason-etal09}, which 
are confirmed by X-ray observations \citep{Allen-etal02,Ota-etal08}. 
The shock velocity inferred from the SZ effect and the X-ray data of
RX~J1347-1145 is 4600~km~s$^{-1}$ \citep{Kitayama-etal04}, which is
similar to the shock velocity observed in 1E0657-56
\citep{Markevitch06}. The lack of structure in the redshift distribution
of member galaxies of RX~J1347-1145 suggests that the geometry of the
merger of this cluster is also closer to edge-on \citep{Lu-etal10}.
However, the lack of a bow shock in the {\sl Chandra} image may suggest 
that it is not quite as edge-on as 1E0657-56. In any case, it seems plausible 
that there may be more clusters like 1E0657-56 in our universe. This too may 
present a challenge to $\Lambda$CDM. 

Since the volume of the MICE simulation is close to the Hubble 
volume,\footnote{For example, the comoving volume available from $z=0$
to $z=1$ over the full sky is $54~h^{-3}~{\rm Gpc}^3$, which is only
twice as large as the volume of the MICE simulation. The
comoving volume out to $z=3$ is still $396~h^{-3}~{\rm Gpc}^3$, which
is nowhere near enough to overcome the probability of $10^{-9}$.} our results can be compared directly with observations, provided
that detailed follow-up observations are available for us to calculate
the shock velocity, gas distribution, and dark matter distribution. 
These three observations would then enable us to estimate the mass ratio and
initial velocity of the collision which, in turn, can be compared to the
probability distribution we have derived in this paper.
Note also that the probabilities obtained in our work are the conditional 
ones. That is, the probability for which a fitting formula is  
provided is the probability of the velocity of bullet-systems that are 
nearly head-on, with 1:10 or more mass ratio, and with  $M_{main}\ge 
0.7\times10^{15}\, h^{-1}M_{\odot}$. If we computed the probability of finding 
high-velocity bullet systems among all clusters from the simulation, then 
the probability would be even smaller than those estimated above. Such a
conditional probability is relevant to the observation, if we have
sufficient amount of data for estimating the mass ratio and the initial
velocity, as mentioned above. Note that we have precisely such data for
1E0657-56. 

An interesting question that we have not addressed in this paper is how
many high-velocity bullet systems are expected for flux-limited galaxy
cluster surveys, such as the South Pole Telescope and eROSITA (extended
ROentgen Survey with an Imaging Telescope Array). To calculate, e.g.,
$dN_{\rm bullet}/dz$, one
needs the light-cone output of the MICE simulation. While we have not
investigated this, we expect two major light-cone effects on the infall
velocity distribution.  
First, the infall velocities at high $z$'s should be larger 
since the effect of $\Lambda$ has yet to kick in at high $z$'s, which we 
have already demonstrated here by comparing the mean infall velocity at 
$z=0$ and $0.5$. Second, the massive bullet systems with mass greater than 
$10^{15}\, h^{-1}M_{\odot}$ are very rare at higher $z$'s. 
The first effect will make  the high-velocity system more common, while the 
second effect will make the high-velocity system less common. In order
to quantify the net effect, one needs the light-cone output.
However, the light-cone effect alone would not be able to
reconcile the existence of 1E0657-56 with the prediction of $\Lambda$CDM.

\acknowledgments

We acknowledge the use of data from the MICE simulations that are 
publicly available at http://www.ice.cat/mice. We thank C.~Mastropietro, 
M.~Milosavljevi\'c and P.~R. Shapiro for discussion. We also thank 
an anonymous referee for helpful comments. 
J.L. is very grateful to the members of the Texas Cosmology Center of
the University of Texas at Austin for the warm hospitality during the
period of her visit when this work was initiated and
performed. J.L. acknowledges the financial support  
from the Korea Science and Engineering Foundation (KOSEF) grant funded by 
the Korean Government (MOST, NO. R01-2007-000-10246-0).
This work is supported in part by a NASA grant NNX08AL43G and an NSF grant
AST-0807649.

\clearpage
\clearpage
\clearpage
\clearpage
\clearpage
\clearpage
\clearpage
\begin{deluxetable}{cccc}
\tablewidth{0pt}
\setlength{\tabcolsep}{5mm}
\tablehead{$z$ & the number of & the number of & the mean mass of \\
& clusters & clusters of clusters\tablenotemark{a} & main clusters}
\startdata
$0$ & 2.8 million & 0.29 million & $1.3\times 10^{14}~h^{-1}~M_\sun$\\  
$0.5$ & 1.7 million & 0.20 million & $1.1\times 10^{14}~h^{-1}~M_\sun$
\enddata
\tablenotetext{a}{A ``cluster of clusters'' is a group of 
cluster-size halos identified by the FoF algorithm. A useful picture is a
massive cluster surrounded by many less massive clusters.}
\label{tab:fof}
\end{deluxetable}
\clearpage
\begin{deluxetable}{cccc}
\tablewidth{0pt}
\setlength{\tabcolsep}{5mm}
\tablehead{$M_{\rm main}$ & the number of &  the number of & the number
 of \\
$[10^{15}h^{-1}M_\sun]$ & clusters of clusters &  bullet-like
 systems\tablenotemark{a} & bullet-like systems\tablenotemark{b}\\
&at $z=0$& at $z=0$& at $z=0$}
\startdata
$\ge 0.5$ & $8523$ & $2189$ & $3093$\\ 
$\ge 0.7$ & $3135$ & $1135$ & $1402$ \\ 
$\ge 1$ & $911$ & $351$ & $391$
\enddata
\tablenotetext{a}{For $M_{\rm sat}/M_{\rm main}\le 1/10$. 
A ``bullet-like system'' is defined as a nearly head-on collision system 
satisfying all of the conditions (1, 2, and 3) given in Section~3.}
\tablenotetext{b}{For $M_{\rm sat}/M_{\rm main}\le 1/5$.}
\label{tab:z0}
\end{deluxetable}
\clearpage
\begin{deluxetable}{cccc}
\tablewidth{0pt}
\setlength{\tabcolsep}{5mm}
\tablehead{$M_{\rm main}$ & the number of  &  the number of & the
 number of \\
$[10^{15}h^{-1}M_\sun]$ & clusters of clusters & bullet-like
 systems\tablenotemark{a} & bullet-like systems\tablenotemark{b}\\
&at $z=0.5$& at $z=0.5$& at $z=0.5$} 
\startdata
$\ge 0.5$ & $3108$ & $186$ & $240$\\
$\ge 0.7$ & $800$ & $78$  & $93$\\
$\ge 1$ & $138$ & $27$ & $32$
\enddata
\tablenotetext{a}{For $M_{\rm sat}/M_{\rm main}\le 1/10$.} 
\tablenotetext{b}{For $M_{\rm sat}/M_{\rm main}\le 1/5$.}
\label{tab:z05}
\end{deluxetable}
\end{document}